\documentclass[aps,prb,twocolumn,preprintnumbers,amsmath,amssymb]{revtex4}
\usepackage{graphicx}
\usepackage{xcolor}
\begin{document}
\title{On the heat capacity of quantum hard sphere fluid}

\author{S.M. Stishov}
\email{stishovsm@lebedev.ru}
\affiliation{P. N. Lebedev Physical Institute, Leninsky pr., 53, 119991 Moscow, Russia}

\begin{abstract}
The thermodynamic properties of the Boltzmann hard sphere system is discussed. It was found that zero point energy decreases with temperature so slowly that it turned out to be an almost a constant addition to the classical value. In result the heat capacity of the system differs little from the classical value of 3/2 k everywhere except for the narrow region of low temperatures, where heat capacity drops to zero. The predicted linear temperature contribution to the heat capacity like in ideal Fermi gas was clearly detected in the quantum hard sphere system at the lowest temperatures.    
\end{abstract}

\maketitle
\section{Introduction} At sufficiently high temperatures, or in systems with a strong repulsive interaction, when the particles exchanges are practically impossible, the effects of Bose and Fermi statistics can be neglected. However, the system may be quantum mechanical due to "diffraction effects" associated with the wave nature of the particles. Moreover, the effects of quantum statistics decay exponentially with increasing temperature, while the "diffraction effects" disappear as an inverse power of temperature at $T\rightarrow\infty$. 
Thus, in the quantum system of hard spheres there is a significant temperature range where the effects of quantum statistics play only a minor role~\cite{runge}. So further we will discuss energy and heat capacity behavior of the Boltzmann quantum hard sphere fluid.
\section{Discussion and results} 
\begin{figure}[htb]
\includegraphics[width=40mm]{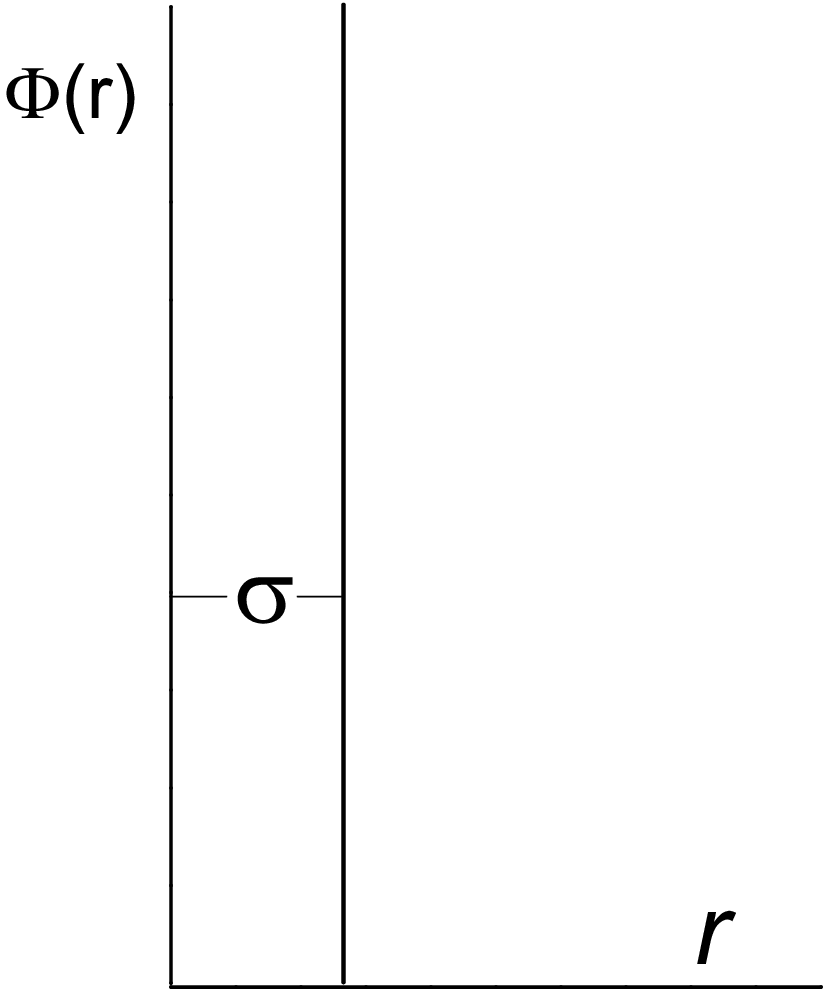}
\caption{\label{fig1} Hard sphere interaction potencial .}
\end{figure}

The system of classical hard spheres is the simplest non-trivial system with an interaction of the form Fig.\ref{fig1}:
\begin{equation}
\label{eq1}
\begin{aligned}
\Phi(r)=0,\ r>\sigma \\    
\Phi(r)= \infty,\ r<\sigma
\end{aligned}
\end{equation}
However, in contrast to the classical system of hard spheres, in the quantum case an interparticle repulsion occurs due to the uncertainty principle, which ensures an existence of the “restoring” force to long-wavelength acoustic deformations~\cite{runge}.
The hard sphere model has been widely used to describe strongly interacting systems. Let us recall the van der Waals theory of critical phenomena, in which the interparticle repulsive interaction is described by the hard sphere potential. Subsequently, much effort was expended developing a theory of fluids using the hard sphere model as a zero approximation in the framework of perturbation theory~\cite{barker}. The quantum model of hard spheres has been used at an analysis of behavior of quantum systems with short-range interactions, in particular, helium~\cite{Hansen,kalos}. 
\begin{figure}[htb]
\includegraphics[width=60mm]{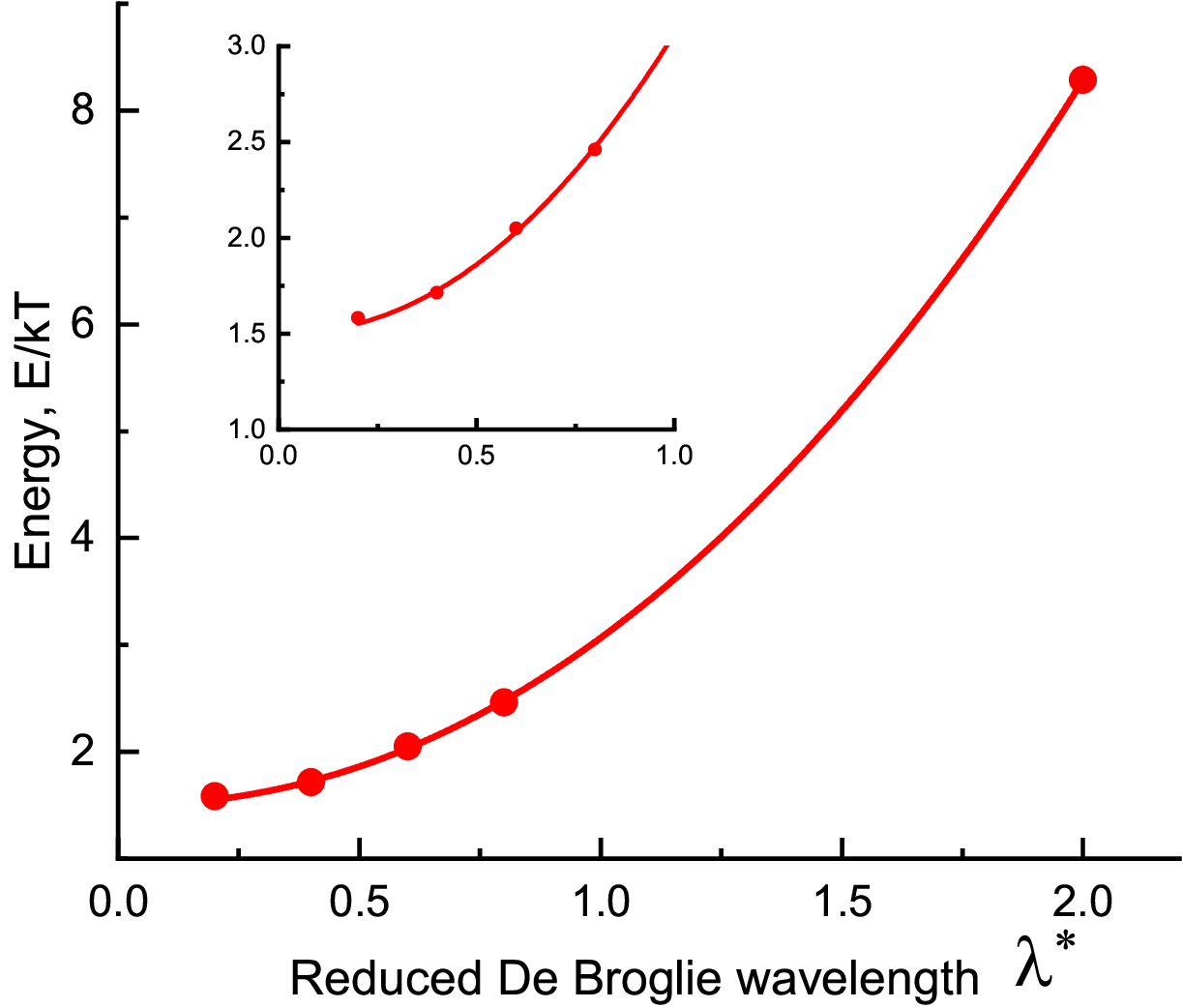}
\caption{\label{fig2} Dependence of the dimensionless energy $E/kT$ of the fluid of quantum hard spheres on the reduced De Broglie wavelength $\lambda^{*}=h/(2\pi mkT\sigma^{2})^{1/2}$, circles - results of calculations~\cite{sese1}, line – approximation.}
\end{figure}

Now we turn attention to one particular study on thermodynamic properties of quantum hard spheres published many years ago in Ref.\cite{Cole}. A surprising result of this study was claim of Fermi-like linear temperature dependence of heat capacity system $C_{v}\backsim T$, which arises from “the physical exclusion of interpenetration rather than statistics”~ \cite{Cole}. But real physics of this situation  can not be described in a simple way.
 
Indeed, in a dense hard sphere system particles are confined in some sort of cage formed by the neighboring particles. Then an energy of the particles should be quantisized. But because of non regular forms of cages in a hard sphere fluid the corresponding energy levels should be different for each particular cage. Curious that the calculations~\cite{Ros} of specific heat of a quantum particle in a box do not show a linear behavior at low temperature.   
A validation of the cited study~\citep{Cole} can be conducted with results of calculations of the thermodynamic properties of the quantum system of hard spheres by the Monte Carlo method, carried out in Ref.~\cite{sese1}. The author ~\cite{sese1} presented the values of the dimensionless energy $E/kT$ of the fluid state of the system as a function of the reduced density $\rho^{*}=\rho\sigma^{3}$ ($\sigma$-sphere diameter) along lines with constant $\lambda^{*}$($\lambda^{*}=h/(2\pi mkT\sigma^{2})^{1/2}$  is the ratio of the thermal de Broglie wavelength to the diameter of the hard sphere).
For analysis, we select the results of energy calculations at density $\rho^{*}=0.3$, covering the largest range of reduced de Broglie lengths $\lambda^{*}$. The corresponding data is shown in Fig.~\ref{fig2}.
As can be seen from Fig.2, the calculated data are obviously extrapolated at $\lambda^{*}\rightarrow 0$ to the classical value of $E/kT=1.5$, which verifies the calculated results. Note that the total energy of quantum hard spheres includes only the kinetic energy of the translational motion of particles and zero energy associated with the uncertainty principle. The approximation formula describing the numerical data~\cite{sese1} has the form:
\begin{figure}[htb]
\includegraphics[width=50mm]{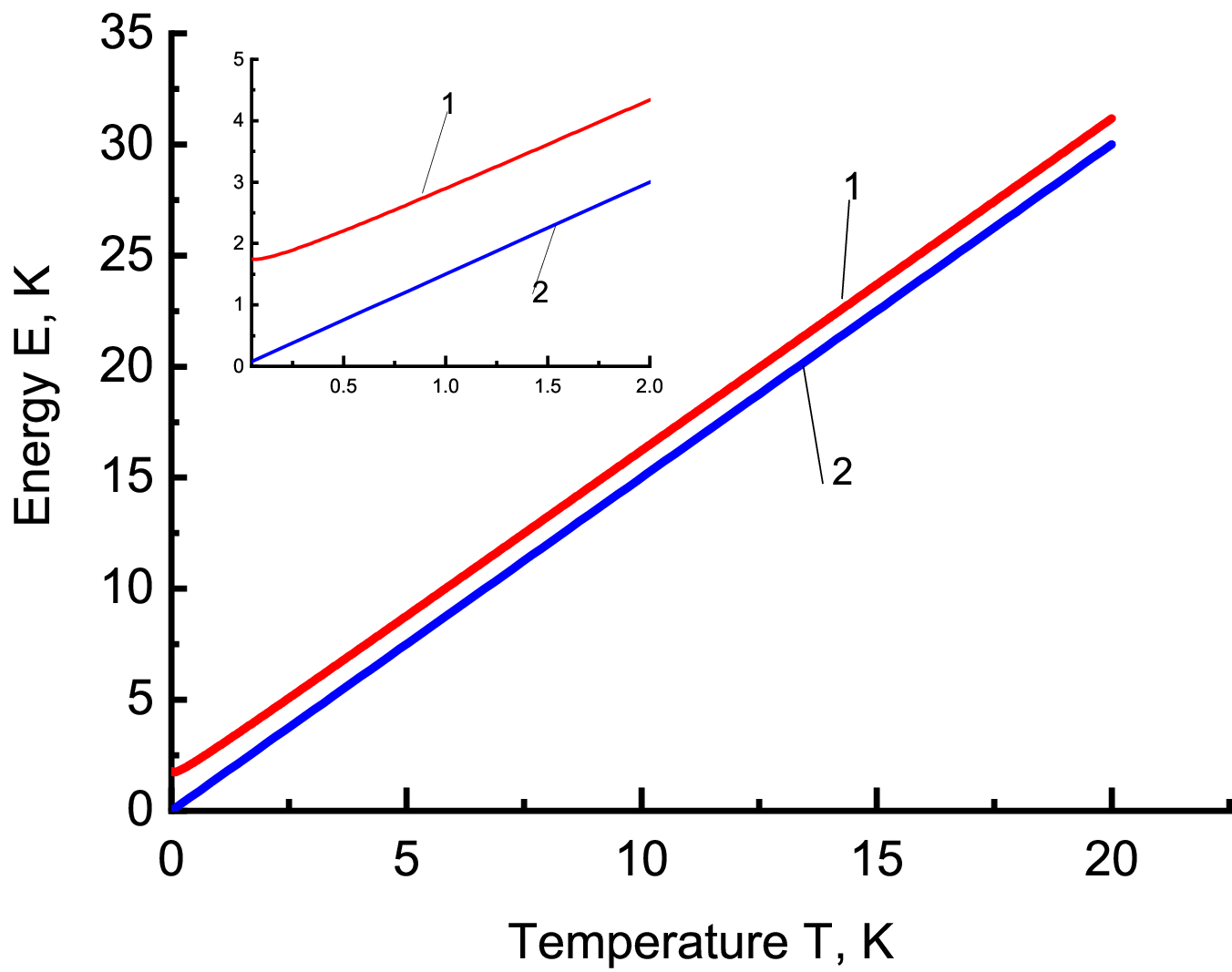}
\caption{\label{fig3} Dependence of the energy of quantum (1) and classical (2) systems of hard spheres on temperature.}
\end{figure}
\begin{equation}
\label{eq2}
\begin{aligned}
E/kT=1.5 + 1.5645({\lambda^*})^{2.1169}
\end{aligned}
\end{equation} 
Substituting the numerical data into the expression $\lambda^{*}$ ( $\sigma=3.5$ \AA, m=28.0134 a.u.~\cite{sese1} )we obtain for the energy and heat capacity:
\begin{equation}
\label{eq3}
\begin{aligned}
E=1.5T+1.395 T^{-0.06}  
\end{aligned}
\end{equation} 
\begin{equation}
\label{eq4}
\begin{aligned}
C_{v}=1.5-0.084T^{-1.006}     
\end{aligned}
\end{equation} 
Note, as follows from a relation (4) $C_{v}$ turns to zero at small but finite temperature equal to $~5.7 \times 10^{-2}$, which is obviously a result of calculational errors and approximations. This mismatch is corrected when needed.

Quite surprising results follow from expressions (3) and (4).
Zero point energy decreases so slowly with temperature that it turns out to be an almost a constant addition to the classical value, Fig.\ref{fig3}. 
The behavior of the quantum contribution to the energy of a system of hard spheres (Fig.~\ref{fig3}) confirms the conclusion of the work~\cite{bha} that as a contrary to naive expectations, quantum effects turn out to be very important even when the thermal wavelength of De Broglie is only a small fraction of the hard sphere diameter.
Due to the mentioned specifics of the quantum contribution, the heat capacity of the system differs little from the classical value of 3/2 k everywhere except for the narrow region of low temperatures, where heat capacity of the system drops to zero (Fig. 4).
\begin{figure}[htb]
\includegraphics[width=60mm]{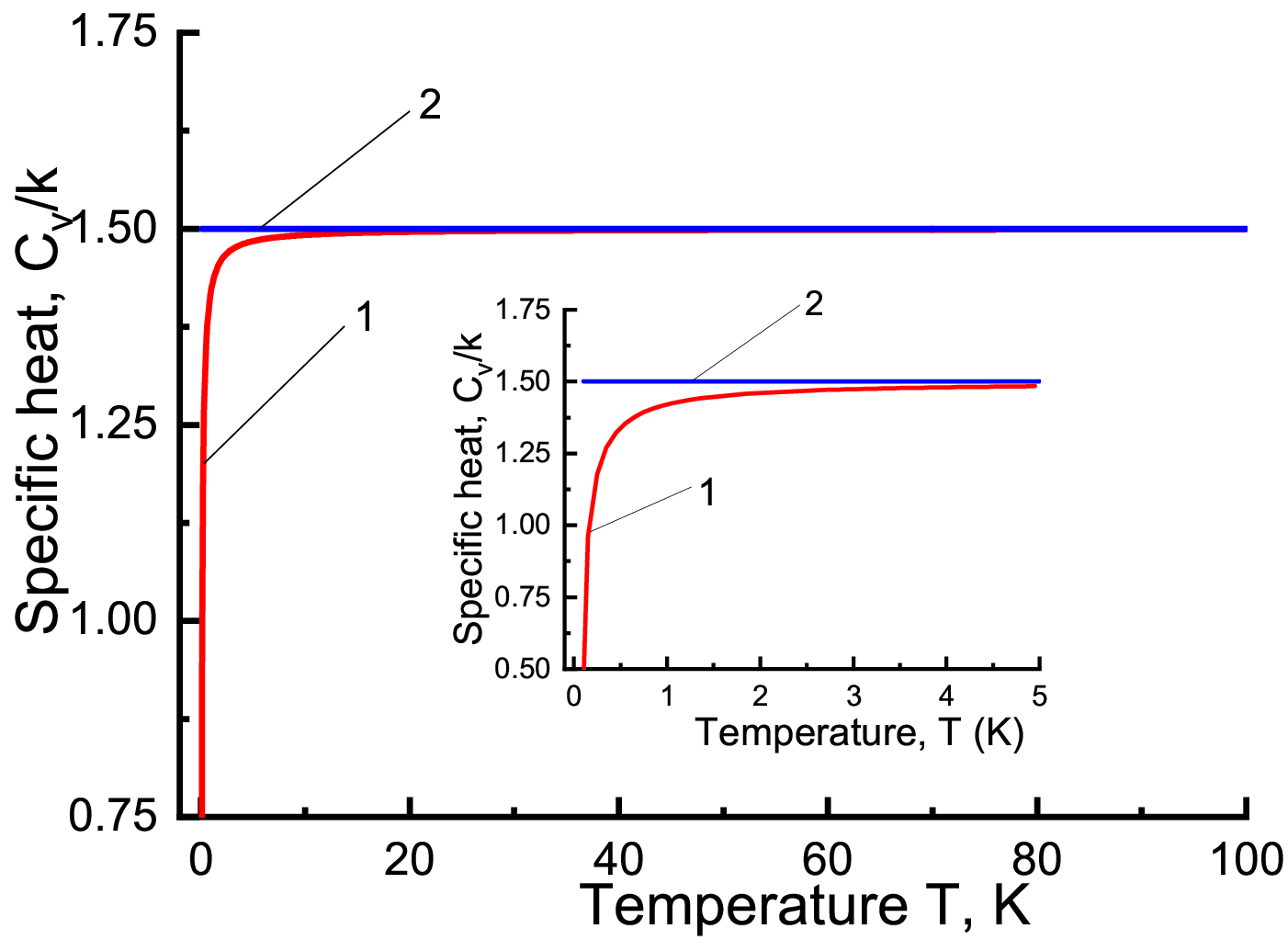}
\caption{\label{fig4} Heat capacity of (1) quantum and (2) classical systems of hard spheres. }
\end{figure}
\begin{figure}[htb]
\includegraphics[width=60mm]{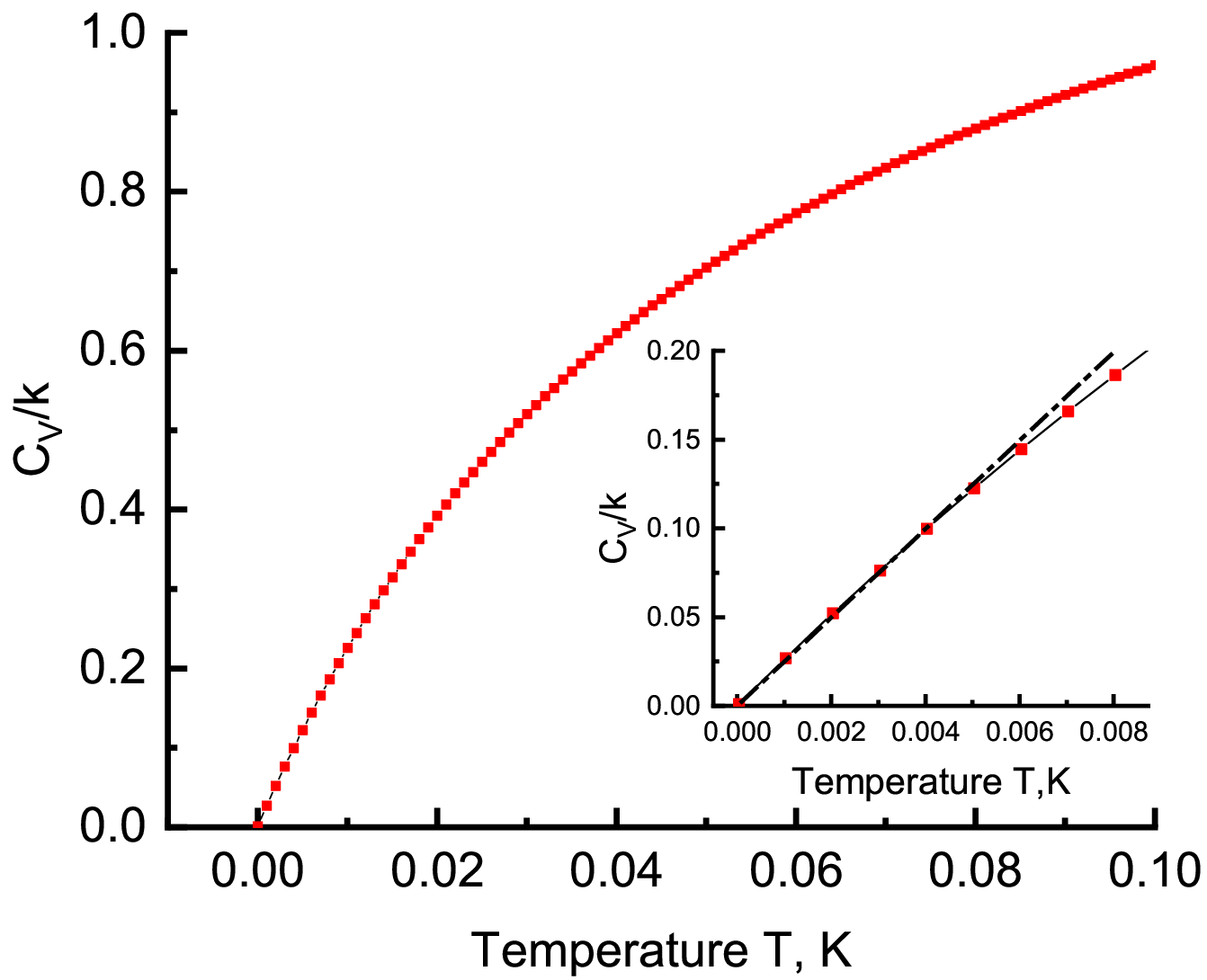}
\caption{\label{fig5} Heat capacity of quantum system of hard spheres at low temperature. }
\end{figure}

\begin{figure}[htb]
\includegraphics[width=60mm]{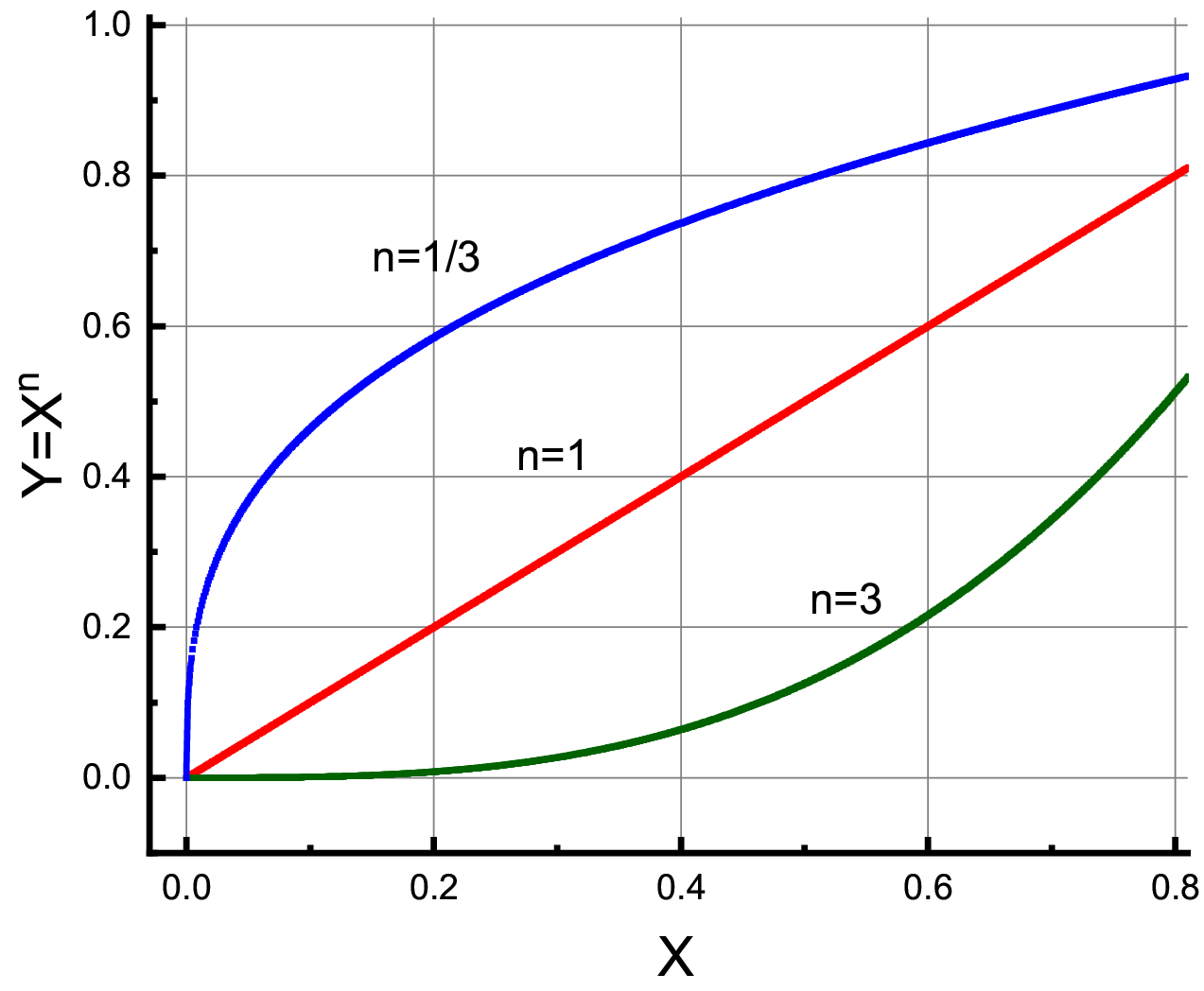}
\caption{\label{fig6} Functions of the form $Y=X^n$ in the vicinity of x=0 . }
\end{figure}

\begin{figure}[htb]
\includegraphics[width=60mm]{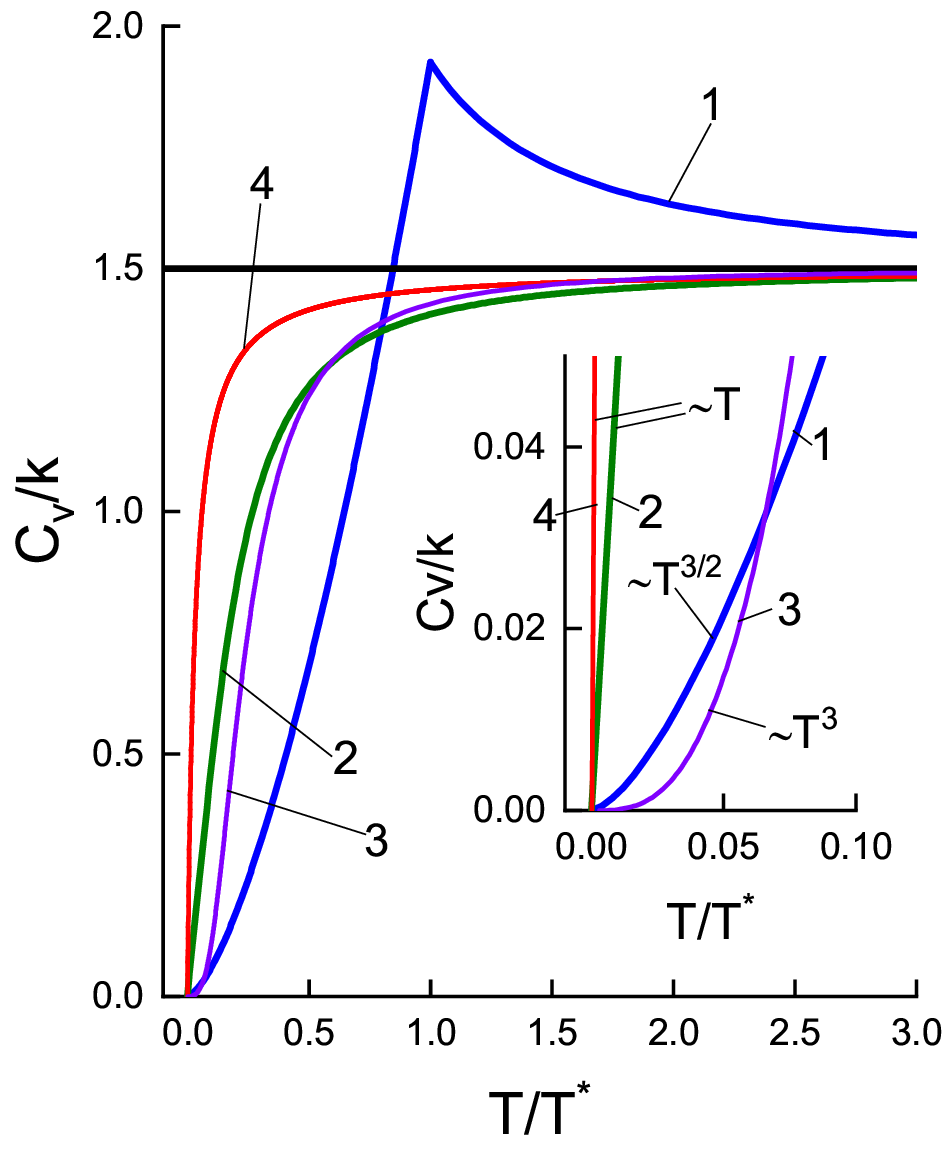}
\caption{\label{fig7}  Normalized heat capacities of model systems (1 - $C_{v}$ of ideal Bose gas, $T^{*}=T_{c}$-phase transition peak, 2 - $C_{v}$ of ideal Fermi gas, $T^{*}=T_{Fermi}$, 3 - $C_{v}$ of Debye solid, $T^{*}=T_{\Theta}$-Debye temperature and 4 - $C_{v}$ of quantum system of hard sphere, $T^{*}=\dfrac{\hbar^{2}}{m\sigma^{2}}=5.7 \times 10^{-1}K$). }
\end{figure}
\begin{figure}[htb]
\includegraphics[width=60mm]{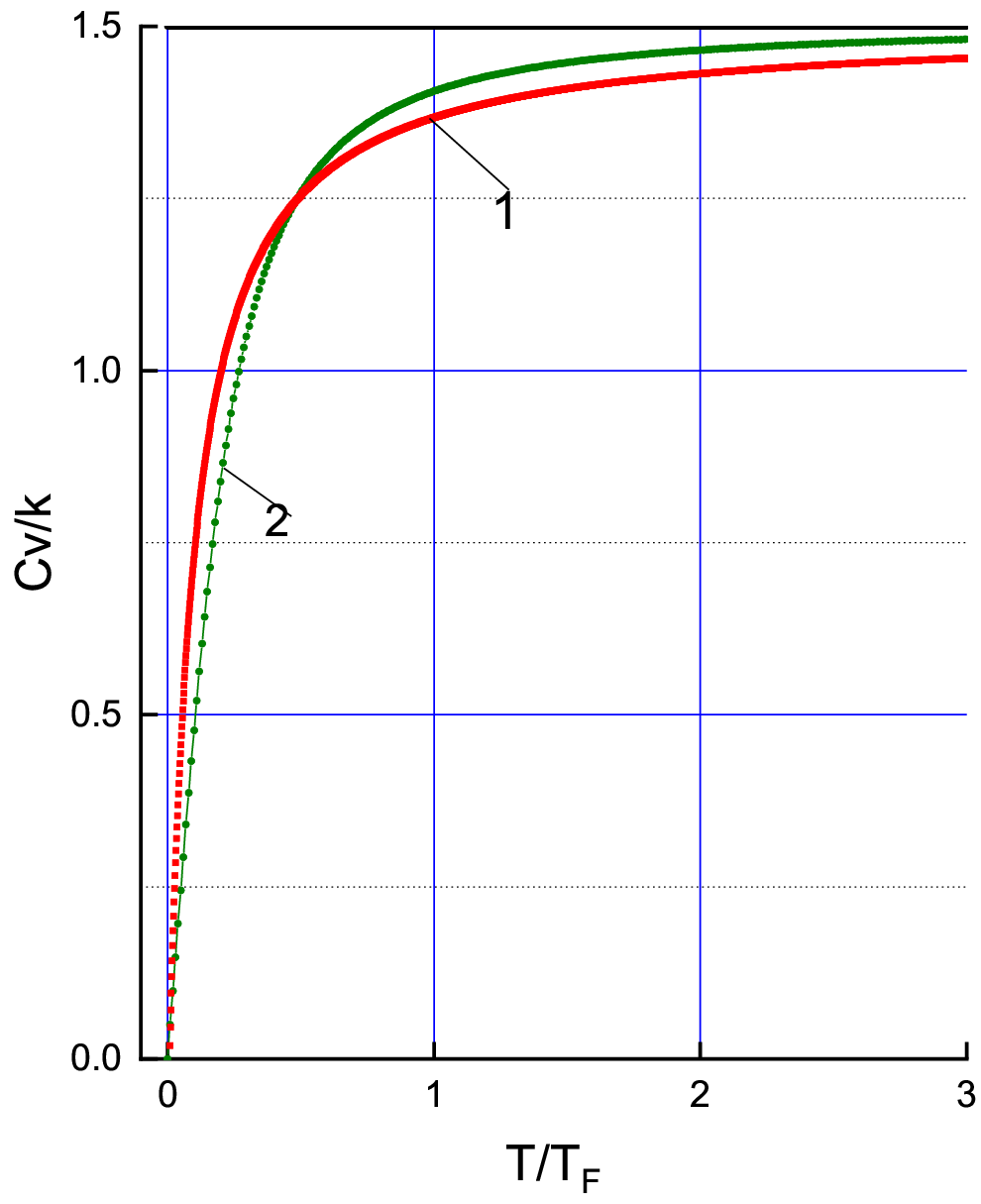}
\caption{\label{fig8} Reduced heat capacity of quantum system of hard spheres (1) comparing to heat capacity of Fermi ideal gas (2). }
\end{figure}
The low temperature part of the heat capacity of quantum system of hard spheres is depicted in Fig.~\ref{fig5}. As is seen the dependence $C_{v}(T)$  certainly contains the low temperature linear component. A finite value of the derivative $dC_{v}/dT$ at the coordinate origin like it occurs in case of the Fermi gas clearly support this conclusion. Fig.~\ref{fig6} well illustrates this point.  It should be reminded that the linear temperature dependence of heat capacity of Fermi gas arises only at $T/T_{f}<<1$, where $T_{f}$-Fermi energy. At higher temperatures a behavior of heat capacity is essentially nonlinear (see Ref.\cite{Pat}). The same situation is expected in our case and a linear behavior of heat capacity can be seen at $T/\varepsilon <<1$, where $\varepsilon$ is some energy barrier, preventing particles from free moving. One may conclude from Fig.\ref{fig5} that $\varepsilon \thickapprox 10^{-2}K$. 
    
In this connection it is instructive to analyze Fig.\ref{fig7}, where four $C_{v}(T)$ curves describing heat capacity behavior as functions of temperature of ideal Bose and Fermi gases, quantum Boltzmann and Debye solid are displayed. One can see in Fig.\ref{fig7} that the curves exhibit different behavior in the vicinity of zero temperature certainly as a result of different nature of excitations responsible for the heat capacity (single particle or collective). 
Probably just single particle character of thermal excitations in the hard sphere fluid and ideal Fermi gas defines their linear dependence of heat capacity on temperature. The distinct similarity of heat capacity curves of ideal Fermi gas and hard sphere fluid is illustrated in Fig.\ref{fig8}.

\section{Conclusion}
Heat capacity behavior as a function of temperature  of the Boltzmann quantum hard sphere fluid was  derived from the Quantum Monte Carlo calculations, which appeared to be quite similar to that of the ideal Fermi gas. We suggest that the reason of this similarity lies in the specifics of single particle nature of excitations responsible for heat capacity characteristics in the both media.    
\section{Acknowlegment} Author appreciates A.M. Belemuk advice on the matter of the Fermi gas properties and expresses his gratitude to A.E. Petrova for some calculations. 
	
\end{document}